\documentclass[reprint,
amsmath,amssymb]{revtex4-1}
\usepackage[utf8x]{inputenc}
\usepackage{graphicx}
\usepackage{dcolumn}
\usepackage{bm}

\begin{document}
\title[Cumulants of Berry phase]{Cumulants associated with geometric phases}
\author{Bal\'azs Het\'enyi and Mohammad Yahyavi}
\address{Department of Physics \\
Bilkent University \\
TR-06800 Bilkent, Ankara, Turkey
}

\date{\today}

\begin{abstract}
The Berry phase can be obtained by taking the continuous limit of a cyclic
product $-\mbox{Im} \ln \prod_{I=0}^{M-1} \langle \Psi_0({\boldsymbol
  \xi}_I)|\Psi_0({\boldsymbol \xi}_{I+1})\rangle$, resulting in the circuit
integral $i \oint \mbox{d}{\boldsymbol \xi} \cdot \langle \Psi_0({\boldsymbol
  \xi})|\nabla_{\boldsymbol \xi}|\Psi_0({\boldsymbol \xi}\rangle$.
Considering a parametrized curve ${\boldsymbol \xi}(\chi)$ we show that the
product $\prod_{I=0}^{M-1} \langle \Psi_0(\chi_I)|\Psi_0( \chi_{I+1})\rangle$
can be equated to a cumulant expansion.  The first contributing term of this
expansion is the Berry phase itself, the other terms are the associated 
spread, skew, kurtosis, etc.  The cumulants are shown to be gauge invariant.
It is also shown that these quantities can be expressed in terms of an
operator.
\end{abstract}

\maketitle

{\it Introduction.}  The concept of geometric phase was first suggested by
Pancharatnam~\cite{Pancharatnam56} in optics.  In 1984 Berry~\cite{Berry84}
published a paper about phases which arise when a quantum system is
brought around an adiabatic cycle.  The phase advocated in this paper was
overlooked earlier~\cite{Fock28} as it was considered part of the arbitrary
phase of a quantum wavefunction.  Berry has shown that this is not the case,
and that the phase of an adiabatic cycle can be a measurable quantity.  Since
the publication of Berry's paper this concept was found to be at the
core~\cite{Shapere89,Xiao10} of a number of interesting physical effects,
including the Aharonov-Bohm effect~\cite{Aharonov59}, quantum Hall
effect~\cite{Thouless82}, topological insulators~\cite{Hasan10}, dc
conductivity~\cite{Hetenyi13}, or the modern theory of
polarization~\cite{King-Smith93,Resta94}.  More recently an example of a
geometric phase, the Zak phase, has been measured in optical
waveguides\cite{Longhi13} and optical lattices~\cite{Atala13}.

To derive a Berry phase, one considers a Hamiltonian which depends
parametrically on a set of variables.  One can then take a discrete set of
points in this parameter space, obtain the wavefunction, and form a cyclic
product of the type in Eq. (\ref{eqn:phi_dscrt}).  The imaginary part of the
logarithm of this cyclic product corresponds to the discrete Berry phase.  If
the discrete points are along a cyclic curve then the continuous limit can be
taken, and it corresponds to the well-known circuit integral~\cite{Berry84}.
The real part of the product is usually not considered, due to the common
belief that as a result of the normalization of the wave-function, it is zero,
therefore not physically relevant.  In this paper we consider a curve
parametrized by a scalar and show that when the real part of the product in
Eq. (\ref{eqn:phi_dscrt}) is considered it leads to a physically well-defined
quantity.  More generally we show that the product in
Eq. (\ref{eqn:phi_dscrt}) can be expanded using the standard cumulant
expression, with the first order term corresponding to the Berry phase, and
the higher order terms giving gauge invariant and therefore physically
well-defined quantities.  It is also shown that the cumulants can be written
in terms of an operator.  We then consider the spread of polarization, which
was given by Resta and Sorella~\cite{Resta99}, and show that the spread
suggested by them coincides with that obtained from the cumulant expansion
described here.  We also analyze one of the canonical examples for the Berry
phase~\cite{Berry84} in light of our findings.

{\it General remarks.}  The most general way to obtain the Berry phase is to
write it in the discrete representation, and then take the continuous limit.
Pancharatnam's~\cite{Pancharatnam56} original derivation is based on
considering discrete phase changes.  The discrete Berry phase first appeared
in 1964, in a paper by Bargmann~\cite{Bargmann64}, as a mathematical tool for
proving a theorem.  The expression which forms the basis of our derivation
here has also been used extensively in the case of path-integral based
representation of geometric phases~\cite{Kuratsuji85,Kuratsuji86}.

Given a parameter space ${\boldsymbol \xi}$ and some Hamiltonian $H({\boldsymbol
  \xi})$ with
\begin{equation}
H({\boldsymbol \xi})| \Psi_i({\boldsymbol \xi})\rangle = E_i({\boldsymbol \xi}) | \Psi_i({\boldsymbol \xi})\rangle,
\end{equation}
where $| \Psi_i({\boldsymbol \xi})\rangle$($E_i({\boldsymbol \xi})$) is an
eigenstate(eigenvalue) of the Hamiltonian.  Consider a set of $M$ points in
this parameter space $\{{\boldsymbol \xi}_I\}$.  In this case one can form the
quantity
\begin{equation}
\label{eqn:phi_dscrt}
\phi = -\mbox{Im} \ln  \prod_{I=0}^{M-1} \langle \Psi_0({\boldsymbol \xi}_I)|\Psi_0({\boldsymbol \xi}_{I+1})\rangle,
\end{equation}
where $\Psi_0({\boldsymbol \xi}_{M})=\Psi_0({\boldsymbol \xi}_0)$ (cyclic)
which is physically well-defined since arbitrary phases cancel.  In
Eq. (\ref{eqn:phi_dscrt}) $\phi$ is formed using the ground state, without
loss of generality.  If the points $\{{\boldsymbol \xi}_I\}$ are points on a
closed curve, one can take the continuous limit and obtain
\begin{equation}
\label{eqn:phi_cont}
\phi = i \oint \mbox{d}{\boldsymbol \xi} \cdot \langle \Psi_0({\boldsymbol
  \xi})|\nabla_{\boldsymbol \xi}|\Psi_0({\boldsymbol \xi}\rangle.
\end{equation}

The $\phi$ can be shown to be gauge invariant and is therefore a physically
well-defined quantity.  If the wavefunction can be taken to be real, then a
nontrivial Berry phase corresponds to $\phi=\pi$ and will only occur if the
enclosed region of parameter space is not simply connected.  If the
wavefunctions can not be taken as real then a non-trivial Berry phase can
occur even if the parameter space is not simply connected.

{\it Cumulant expansion of the Bargmann invariant.}  We consider the product
in Eq. (\ref{eqn:phi_dscrt}) along a cyclic curve.  We assume that the curve
is parametrized according to a scalar hence the product is $\prod_{I=0}^{M-1}
\langle \Psi_0(\chi_I)|\Psi_0(\chi_{I+1})\rangle$.  We also assume that the length
of the curve is $\Lambda$ and that $\chi_I$ defines an evenly spaced (spacing
$\Delta \chi$) grid.  We start by equating this product to a cumulant
expansion,
\begin{equation}
\label{eqn:cum_exp} 
\prod_{I=0}^{M-1} \langle \Psi_0(\chi_I)|\Psi_0(\chi_{I+1})\rangle = 
\exp\left( \frac{1}{\Delta \chi} \sum_{n=1}^\infty \frac{(i \Delta \chi)^n}{n!}C_n\right).
\end{equation}
Straightforward algebra and taking the continuous limit ($\Delta \chi
\rightarrow 0$, $M\rightarrow \infty$, $\Lambda$ fixed) gives 
\begin{eqnarray}
\label{eqn:cmlnts}
C_1 &=& -i \int_0^\Lambda d\chi \gamma_1 \\ \nonumber 
C_2 &=& - \int_0^\Lambda d \chi [\gamma_2 - \gamma_1^2] \\ \nonumber
C_3 &=& i \int_0^\Lambda d \chi [\gamma_3 -3 \gamma_2 \gamma_1+ 2\gamma_1^3] \\ \nonumber 
C_4 &=& \int_0^\Lambda d \chi [\gamma_4 -3 \gamma_2^2
  -4\gamma_3\gamma_1 + 12 \gamma_1^2\gamma_2 -6\gamma_1^4]
\end{eqnarray}
with $\gamma_i = \langle \Psi_0(\chi) | \partial^i_\chi | \Psi_0(\chi)\rangle$.

$C_1$ corresponds to the Berry phase.  The other $C_i$ look very similar to
the usual cumulants (compare coefficients), provided that we can interpret
$-i\partial_\chi$ as an operator and the integral as an expectation value.
$C_1$ is known to be gauge invariant, therefore it is natural to ask whether
the other $C_i$ are also gauge invariant.  We consider the proof of gauge
invariance for $C_1$.  One first alters the phase of the wavefunction,
i.e. define
\begin{equation}
|\tilde{\Psi}_0(\chi)\rangle = \exp[i \beta(\chi)]|\Psi_0(\chi)\rangle.
\end{equation}
Defining 
\begin{equation}
\tilde{C}_1 = -i\int_0^\Lambda d\chi \langle\tilde{\Psi}_0
(\chi)| \partial_\chi
|\tilde{\Psi}_0(\chi)\rangle,
\end{equation}
it is easy to show that
\begin{equation}
\tilde{C}_1 - C_1 = \beta(\Lambda) - \beta(0).
\end{equation}
with $\tilde{\gamma}_1=\langle \tilde{\Psi_0}(\chi) | \partial^i_\chi |
\tilde{\Psi_0}(\chi)\rangle$.  Hence the Berry phase of the original
wavefunction differs from the shifted one by the difference of
$\beta(\Lambda) - \beta(0)$ which for an adiabatic cycle is $2\pi m$, with
$m$ integer.  Applying the same procedure to the other cumulants we obtain the
following results:
\begin{eqnarray}
\label{eqn:cmlnts_gi}
\tilde{C}_2 - C_2&=&  \dot{\beta}(\Lambda) - \dot{\beta}(0) = 0,\\ \nonumber 
\tilde{C}_3 - C_3&=&  \ddot{\beta}(\Lambda) - \ddot{\beta}(0) = 0,\\ \nonumber 
\tilde{C}_4 - C_4&=&  \dddot{\beta}(\Lambda) - \dddot{\beta}(0) = 0,
\end{eqnarray}
hence, if the function $\beta(\chi)$ and its derivatives are continuous at the
boundaries gauge invariance holds. We have carried out this proof up to fourth
order.  There appears to be a pattern in Eq. (\ref{eqn:cmlnts_gi}).

The cumulants derived above can be expressed in terms of expectation values of
operators.  Consider the expression from perturbation theory
\begin{equation}
\label{eqn:Pert}
\partial_\chi | \Psi_0(\chi) \rangle = \sum_{j\neq0} 
|\Psi_j(\chi)\rangle \langle \Psi_j(\chi)| \frac{\partial_\chi H(\chi) }{
  E_j-E_0} |\Psi_0(\chi) \rangle.
\end{equation}
Defining operator $\hat{O}$ as
\begin{equation}
\partial_\chi H(\chi) = i[H(\chi),\hat{O}]
\end{equation}
it can be shown that the cumulants of this operator correspond to the $C_i$
derived above, except for the case $i=1$, the Berry phase itself, for which
application of Eq. (\ref{eqn:Pert}) leads to zero.  For the Berry phase the
expression from perturbation theory (Eq. (\ref{eqn:Pert} )) is not valid since
it makes a definite choice about the phase of the wavefunction for all values
of $\chi$.  The most general expression is
\begin{eqnarray}
 |\Psi(\chi +\Delta \chi)\rangle = e^{i\alpha}\times \hspace{3cm}\\
\left[|\Psi(\chi)\rangle + 
\sum_{j\neq0} 
|\Psi_j(\chi)\rangle \langle \Psi_j(\chi)| \frac{\partial_\chi H(\chi) }{
  E_j-E_0} |\Psi_0(\chi) \rangle\right], \nonumber
\end{eqnarray}
but in standard perturbation theory $\alpha$ is assumed to be zero.  This
phase difference shifts the first cumulant (the Berry phase), however since it
is a mere shift, it leaves the other cumulants unaffected.  One can conclude
that while the Berry phase itself can not be expressed in terms of an
operator, its associated cumulants can.  This statement will be clarified in
an example below.

{\it Polarization, current and their spreads.}  We now consider the Berry
phase corresponding to the polarization from the modern
theory~\cite{King-Smith93,Resta94,Resta99,Resta98}.  In this theory an
expression for the spread of a Berry phase associated quantity has been
suggested, and we now show that it is equivalent to $C_2/\Lambda$.

Resta showed that the expectation value of the position over some wavefunction
$|\Psi_0\rangle$ of a system with unit cell dimension $L$ can be written as
\begin{equation}
\label{eqn:X}
\langle X \rangle = -\frac{1}{\Delta K} \mbox{Im} \ln \langle \Psi_0 | e^{-i\Delta
  K \hat{X}} | \Psi_0 \rangle,
\end{equation}
where $\Delta K = 2\pi/(N_k L)$, $N_k$ denotes an integer, $\hat{X} = \sum_j
\hat{x}_j$ is the sum of the positions of all particles.  The spread in
position ($\sigma^2_X = \langle X^2\rangle - \langle X \rangle^2$)
can be written
\begin{equation}
\sigma^2_X = -\frac{2}{\Delta K^2} \mbox{Re} \ln \langle \Psi_0 | e^{-i\Delta K
  \hat{X}} | \Psi_0 \rangle,
\end{equation}
The operator $e^{i\Delta K \hat{X}}$ is the total momentum shift operator
which, as has been shown elsewhere~\cite{Hetenyi09,Essler05} has the property
that for a state $|\Psi_0(K)\rangle$ with particular crystal momentum $K$
defined as
\begin{equation}
\Psi_0(k_1 + K,k_2+K,...),
\end{equation}
it holds that 
\begin{equation}
e^{-i\Delta K \hat{X}}|\Psi_0(K)\rangle = |\Psi_0(K+\Delta K)\rangle, 
\end{equation}
in other words it shifts the crystal momentum by $\Delta K$.
To use the shift operator we first write
\begin{equation}
\sigma^2_X  = -\frac{2}{N_k \Delta K^2} \mbox{Re} \ln \langle \Psi_0 |
e^{-i\Delta K \hat{X}} | \Psi_0 \rangle^{N_k}.
\end{equation}
We associate the state $|\Psi_0\rangle$ with a particular crystal momentum
$K_0$,
\begin{equation}
|\Psi_0\rangle = |\Psi_0(K_0)\rangle.
\end{equation}
Using the total momentum shift the scalar product can be rewritten as
\begin{eqnarray}
\nonumber
\langle \Psi_0(K_0) | e^{-i\Delta K \hat{X}} | \Psi_0(K_0) \rangle
&=& \langle \Psi_0(K_0) |  \Psi_0(K_1) \rangle \\ &=& \langle \Psi_0(K_I) |
\Psi_0(K_{I+1}) \rangle,
\label{eqn:TMS}
\end{eqnarray}
where $K_{I+1} = K_I + \Delta K$.  To show the last equation one applies the
Hermitian conjugate of the total momentum shift to $\langle \Psi_0(K_0)|$ $I$
times and the total momentum shift operator to $|\Psi_0(K_0)\rangle$ $I+1$ times
and forms the scalar product.  Thus we can also write
\begin{equation}
\langle \Psi_0(K_0) | e^{-i\Delta K \hat{X}} | \Psi_0(K_0) \rangle^{N_k}
= \prod_{I=0}^{N_k-1} \langle \Psi_0(K_I) |  \Psi_0(K_{I+1}) \rangle.
\end{equation}
The points $K_I$ form an evenly spaced grid with spacing $\Delta K$ in the
Brillouin zone.  Using this result the spread can be rewritten as
\begin{equation}
\sigma^2_X  = -\frac{2}{N_k\Delta K^2} \sum_{I=0}^{N_k}\mbox{Re} \ln 
 \langle \Psi_0(K_I) | \Psi_0(K_{I+1}) \rangle,
\end{equation}
We now expand the scalar product up to second order as
\begin{eqnarray}
\nonumber
\langle \Psi_0(K_I) |  \Psi_0(K_{I+1}) \rangle = 1 + 
\Delta K \langle \Psi_0(K_I) |\partial_K|  \Psi_0(K_I) \rangle + \\
\frac{\Delta K^2}{2} \langle \Psi_0(K_I) |\partial^2_K|  \Psi_0(K_I) \rangle. \hspace{1cm}
\end{eqnarray}
Subsequent expansion of the logarithm and keeping all terms up to second order
in $\Delta K$ results in a first order term of the form
\begin{equation}
\frac{N_kL^2}{2\pi^2}\mbox{Re} 
\sum_{I=0}^{N_k-1} \Delta K \langle \Psi_0(K_I) |  \partial_K   |\Psi_0(K_I) \rangle.
\end{equation}
In the continuum limit ($N_k\rightarrow\infty$) the sum turns into the
integral which gives the standard Berry phase, but since this integral is
purely imaginary it will not contribute to the spread.  The final result for
the spread is
\begin{equation}
\label{eqn:X2}
\sigma^2_X  = \frac{L}{2\pi} 
\sum_{I=0}^{N_k-1} \Delta K \sigma^2_X(K_I)
 = 
\frac{L}{2\pi} 
\int_{-\pi/L}^{\pi/L} \mbox{d} K \sigma^2_X(K),
\end{equation}
where
\begin{equation}
\label{eqn:X2k}
\sigma^2_X(K) = -\langle\Psi_0(K)|\partial^2_K|\Psi_0(K)\rangle
+ \langle\Psi_0(K)|\partial_K|\Psi_0(K)\rangle^2.
\end{equation}
Eq. (\ref{eqn:X2}) is actually the average of the spread over the Brillouin
zone.  One can think of $i\partial_K$ as a ``heuristic position
operator''~\cite{Vanderbilt06}, and the quantity $\sigma^2_X(K)$ as the spread
for a wavefunction with crystal momentum $K$.  This spread of the position
operator, derived by different means, has also been obtained by Marzari and
Vanderbilt~\cite{Marzari97}.  One can also start from the expression for the
spread of the total current~\cite{Hetenyi12b}
\begin{equation}
\sigma^2_K = -\frac{2}{\Delta X^2} \mbox{Re} \ln \langle \Psi_0 | e^{-i\Delta X
  \hat{K}} | \Psi_0 \rangle,
\end{equation}
and apply exactly the same steps as in the case of the total position.  This
derivation results in
\begin{equation}
\sigma^2_K  =
-\frac{1}{L} 
\int_0^L \mbox{d} X 
[\langle\Psi_0(X)|\partial^2_X|\Psi_0(X)\rangle
-
\langle\Psi_0(X)|\partial_X|\Psi_0(X)\rangle^2].
\label{eqn:K2}
\end{equation}

\begin{figure}[ht]
 \centering
 \includegraphics[width=6cm,keepaspectratio=true]{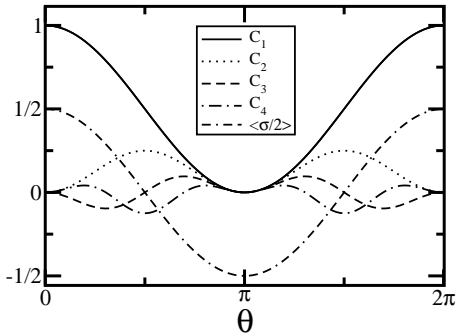}
 \caption{Cumulants of a spin$-\frac{1}{2}$ particle in a precessing field.}
 \label{fig:Cs}
\end{figure}

{\it Example.}  We now calculate the cumulants up to fourth order for one of
the canonical examples for the Berry phase~\cite{Berry84}, a
spin-$\frac{1}{2}$ particle in a precessing magnetic field.  The Hamiltonian
is given by
\begin{equation}
\hat{H}(t) = -\mu {\bf B}(t)\cdot {\boldsymbol \sigma},
\end{equation}
where ${\boldsymbol \sigma}$ are the Pauli matrices, and ${\bf B}(t)$ denotes the
magnetic field,
\begin{equation}
{\bf B}(t) = \left[ \begin{array}{c} \sin\theta \cos \phi \\ \sin\theta \sin \phi \\ \cos\theta\end{array} \right].
\end{equation}
The $z$-component of the field is fixed, the projection on the $x-y$-plane is
performing rotation, i.e. $\phi=\omega t$.  We can proceed to evaluate the
Berry phase and the associated cumulants by defining an adiabatic cycle in
which $\phi$ rotates from zero to $2\pi$.  Using one of the eigenstates
\begin{equation}
|n_-(t)\rangle = \left[ \begin{array}{c} -\sin\left(\frac{\theta}{2}\right)
    \\ 
e^{i\phi}\cos\left(\frac{\theta}{2}\right)
 \end{array} \right].
\end{equation}
The associated cumulants (divided by $2\pi$) evaluate to
\begin{eqnarray}
\label{eqn:Cs}
C_1 =&  \cos^2 \left( \frac{\theta}{2}\right), \\
\nonumber
C_2 =& \left[
\cos^2 \left( \frac{\theta}{2}\right) -
\cos^4 \left( \frac{\theta}{2}\right) 
\right], \\
\nonumber
C_3 =& \left[
\cos^2 \left( \frac{\theta}{2}\right) -
3 \cos^4 \left( \frac{\theta}{2}\right) +
2 \cos^6 \left( \frac{\theta}{2}\right)
\right], \\
\nonumber
C_4 =& 
       \left[\cos^2 \left( \frac{\theta}{2}\right) -
7 \cos^4 \left( \frac{\theta}{2}\right) + 
12 \cos^6 \left( \frac{\theta}{2}\right) -
6 \cos^8 \left( \frac{\theta}{2}\right)
\right]. 
\end{eqnarray}
Fig. \ref{fig:Cs} shows the cumulants as a function of the angle $\theta$.
$C_1$, the Berry phase associated with a spin$-\frac{1}{2}$ particle in a
precessing magnetic field is a well-known result.  The spread is zero when the
Berry phase is zero or $\pi$.  The skew changes sign halfway between zero and
$\pi$ and the kurtosis also varies in sign as a function of the angle
$\theta$.

The operator $\hat{O}$ for this example can easily be shown to be the Pauli
matrix $\frac{\sigma_z}{2}$.  The first order cumulant is given by
\begin{equation}
\left \langle \frac{\sigma_z}{2} \right \rangle =  \sin^2 \left( \frac{\theta}{2}\right) - 
\cos^2 \left( \frac{\theta}{2}\right).
\end{equation}
in other words it is merely shifted compared to the Berry phase.  The higher
order cumulants are identical to those in Eqs. (\ref{eqn:Cs}).  In the
operator representation of the Berry phase the meaning of the first and second
cumulants is rendered more clear.  For the value of $\theta$ for which
$\langle \sigma_z /2 \rangle$ is either $\pm\frac{1}{2}$ the spread is zero.
Indeed those are the maximum and minimum values the operator $\sigma_z$ can
take, hence the spread must be zero.  It is obvious from these results that
the cumulants derived from the Bargmann invariant give information about the
probability distribution of the operator associated with the Berry phase. \\

{\it Measurement of $C_i$s.}  While it has been shown that $C_i$ are
physically well-defined their measurement may not be trivial.  The operator
may not exist or be easily written down.  In this case one can proceed as
follows.  Define:
\begin{eqnarray}
{\boldsymbol \Pi} &=& \prod_{I=0}^{M-1} \langle \Psi_0(\chi_I)|  \Psi_0(\chi_{I+1})\rangle, \\
\nonumber
{\boldsymbol \Pi}^{(o)} &=& \prod_{I=0}^{M/2-1} \langle \Psi_0(\chi_{2I+1})|
\Psi_0(\chi_{2I+3})\rangle, \\
\nonumber
{\boldsymbol \Pi}^{(e)} &=& \prod_{I=0}^{M/2-1} \langle \Psi_0(\chi_{2I})|
\Psi_0(\chi_{2I+2})\rangle.
\end{eqnarray}
Using these definitions one can show that
\begin{eqnarray}
C_3 &\approx& \frac{2}{\Delta \chi^2}
\mbox{Im} \ln \left[ 
\frac{
({\boldsymbol \Pi}^{(o)}{\boldsymbol \Pi}^{(e)})^{\frac{1}{2}}
}{{\boldsymbol \Pi}}
\right]
 + O(\Delta \chi^3), \\
C_4 &\approx& \frac{4}{\Delta \chi^3}
\mbox{Re} \ln \left[ 
\frac{
({\boldsymbol \Pi}^{(o)}{\boldsymbol \Pi}^{(e)})^{\frac{1}{4}}
}{{\boldsymbol \Pi}}
\right]
 + O(\Delta \chi^3). \nonumber
\end{eqnarray}

{\it Conclusions.}  In this paper it was shown that there exists a cumulant
expansion associated with the Berry phase.  The starting point was the
Bargmann invariant, which gives rise to the discrete Berry phase.  The
Bargmann invariant was expressed in terms of a cumulant expansion, the first
term of which was shown to correspond to the Berry phase.  Up to fourth order
it was demonstrated that the cumulants are gauge invariant.  It was also shown
that the cumulants derived can also be related to corresponding expectation
values of a particular operator.  Since, in the modern theory of polarization,
an expression for the second cumulant (spread or variance) is already in use,
as a consistency check, equivalence between that and the spread resulting from
the cumulant expansion presented here was shown.  The cumulants were
calculated for a simple example.

{\it Acknowledgments.}  BH acknowledges a grant from the Turkish
agency for basic research (T\"UBITAK, grant no. 112T176).

\end{document}